\documentclass[final]{raa}            

\usepackage{graphicx,times}             
\usepackage{natbib}
\usepackage{amssymb,amsmath}
\bibpunct{(}{)}{;}{a}{}{,}

\usepackage[a4paper=true,dvipdfm=true,pagebackref=true]{hyperref}
\hypersetup{colorlinks = true, linkcolor = green, anchorcolor = red, citecolor = blue, filecolor = red, pagecolor = red, urlcolor = red}

\begin{document}

   \title{Local galaxies with compact cores as the possible descendants of massive compact quiescent galaxies at high redshift
}

   \volnopage{Vol.0 (20xx) No.0, 000--000}      
   \setcounter{page}{1}          

   \author{Ying Gao
      \inst{1}
   \and Lulu Fan
      \inst{2,3,1}
   }

   \institute{Shandong Key Laboratory of Optical Astronomy and Solar-Terrestrial Environment, School of Space Science and Physics, Institute of Space Sciences, Shandong University,  Weihai, Shandong, 264209, China; 
   \\
        \and
             CAS Key Laboratory for Research in Galaxies and Cosmology, Department of Astronomy, University of Science and Technology of China, Hefei 230026, China;  {\it llfan@ustc.edu.cn}\\
        \and
            School of Astronomy and Space Sciences, University of Science and Technology of China, Hefei, Anhui 230026, People's Republic of China; \\
\vs\no
   {\small Received~~20xx month day; accepted~~20xx~~month day}}

\abstract{ In order to test a possible evolutionary scenario of high-$z$ compact quiescent galaxies (cQGs) that they can survive as local compact cores embedded in local massive galaxies with different morphology classes, we explore the star formation histories of local compact cores according to their spectral analysis. We build a sample of 182 massive galaxies with compact cores (${M}_{*,{\rm core}} > 10^{10.6} {\rm M}_\odot$) at $0.02 \leq z \leq 0.06$ from SDSS DR7 spectroscopic catalogue. {\sf STARLIGHT} package is used to analyze the median stacked spectra and derive the stellar ages and metallicities. Our main results show that local compact cores have the average age of about $12.1\pm0.6$ Gyr, indicating their early formation at $z > 3$, which is consistent with the formation redshifts of cQGs at $1<z<3$. Together with previous studies, our result that local compact cores have similar formation redshifts as those of high-$z$ cQGs, supports that local massive galaxies with compact cores are possible descendants of cQGs. Morphological study of local galaxies with compact cores suggests that there would be multiple possible evolutionary paths for high-$z$ cQGs: most of them ($> 80\%$) will evolve into local massive ETGs according to dry minor merger, while some of them ($\sim 15\%$) will build a substantial stellar/gas discs according to the late-time gas accretion and sustaining star formation, and finally grow up to spiral galaxies.
}

   \authorrunning{Y. Gao \& L. Fan }            
   \titlerunning{Local compact cores as the possible descendants of cQGs}  

   \maketitle

%
%
\section{Introduction}\label{sec:intro}

In the local universe, early-type galaxies (ETGs) are the most massive objects with predominantly old stellar population, indicating that most of their stars were formed at redshift $z\geq1.5$ \citep{Renzini2006}. Among a lot of observational scaling relations, ETGs have a remarkably tight luminosity–size/mass–size correlation \citep{Shen2003,Cappellari2013}. However, massive quiescent galaxies (QGs) at high redshift, which have been considered as the progenitors of local massive ETGs, have been found to deviate from the local mass-size relation, being significantly more compact, by a factor of $3-5$, than local ETGs endowed with the same stellar mass \citep{Daddi2005, Trujillo2006,vD2008,Ryan2012,vandeSande2013,Fan2013a,Fan2013b,Belli2014,Belli2017,vanderWel2014,Straatman2015}.

The dramatic size difference between local ETGs and high-$z$ QGs provides an important observational constraint on various models of massive galaxy formation and evolution. The most popular viewpoint is that the compact high-$z$ QGs experience a dramatic galaxy size evolution which increase their sizes to catch-up the local mass-size relation. Several physical mechanisms have been proposed to explain such an evolution, including dry major merger \citep{Boylan-Kolchin2006}, dry minor merger \citep{Hopkins2009a,Naab2009,Oser2010,Trujillo2011} and AGN/supernova feedback \citep{Fan2008,Fan2010,Damjanov2009}. However, none of these mechanisms alone can explain the observed size evolution \citep{Hopkins2010,Trujillo2013}. The predicted size increase by dry major merger is proportional to the stellar mass increase, resulting in moving galaxies along the direction paralleling the local mass-size relation rather than towards it \citep{Boylan-Kolchin2006,Nipoti2009}. Puff-up due to the massive expulsion of gas by the effect of AGN/supernova feedback predicts a fast size evolution \citep{Ragone-Figueroa2011}, making it difficult in explaining the old compact QGs ($>1$ Gyr). Despite both the cosmological simulations and the observational evidence favor dry minor merger as the main driver of size evolution, there are several works suggesting that it can not be the only mechanism. For instance, there are not enough satellites observed around massive galaxies for dry minor mergers \citep{Marmol-Queralto2012} and the size increase is inefficient even assuming a short merger timescale at $z>1$ \citep{Newman2012}. 

In additional to a genuine growth of galaxy size, the observed size difference between local ETGs and high-$z$ QGs has been explained mainly due to the observational effect, termed progenitor bias \citep{Carollo2013,Poggianti2013,Fagioli2016}. In this viewpoint, the arrival of the newly quenched galaxies, which are typically larger than early quenched galaxies, will move the average sizes of ETGs towards the local mass-size relation. However, there is no size difference between young and old QGs at $z\sim1.5$ \citep{Zanella2016}, which is inconsistent with the expectation of progenitor bias. Based on the dynamical studies and the reconstruction of star formation  histories of $z>1$ QGs, the progenitor bias can contribute about half the increase in average sizes of QGs \citep{Belli2014,Belli2015}.

All aforementioned works follow the same assumption that the high-$z$ compact QGs (cQGs) will evolve into the local ETGs. An inside-out growth of the high-$z$ cQGs has been proposed, i.e., compact cores being formed at high $z$ and evolving into the local ETGs by adding the extended stellar envelope according to dry minor mergers \citep{Patel2013,Huang2013}. It is indeed an important and likely major channel for size evolution of the high-$z$ cQGs, but is not necessarily the only one. Another potential evolutionary scenario leading from the high-$z$ cQGs to the compact cores of local ETGs, lenticular galaxies (S0) and even late-type galaxies (LTGs) has been proposed in recent years \citep{Dullo2013,Graham2013,Graham2015}. To test this new evolutionary scenario, \citet{delaRosa2016} compared the number density, photometric and dynamic mass-size and mass-density relations between local massive galaxies with compact cores and the high-$z$ cQGs. Their results were consistent with the high-$z$ cQGs becoming cores of not only local ETGs and S0, but also some LTGs. The latter was consistent with cosmological simulations, which showed that extended star-forming discs can develop around the high-$z$ cQGs \citep{Zolotov2015,Tacchella2016}.    

In this paper, we aim to investigate the connection between the high-$z$ cQGs and the local massive galaxies with compact cores. We explore the star formation histories (SFHs) of compact cores and compare with the high-$z$ cQGs. This paper is organised as follows. In Section \ref{sec:DATA}, we describe the sample selection of local massive galaxies with and without compact cores and the high-$z$ cQGs . In Section \ref{sec:method}, we show our method of stellar population analysis. We show our main results and discussion in Section \ref{sec:result}. In Section \ref{sec:sum}, we summarize our conclusions. Throughout this paper we assume a standard, flat ${\rm \Lambda}$CDM cosmology \citep{Komatsu2011}, with $H_0 = 70$ km~s$^{-1}$, $\Omega_M = 0.3$, and $\Omega_\Lambda = 0.7$.

\section{DATA AND SAMPLE SELECTION} \label{sec:DATA}

\subsection{Data description}

In order to select the local galaxy sample with and without {\it compact cores}, we utilize Sloan Digital Sky Survey (SDSS) Data Release 7 spectroscopic data set \citep{Abazajian2009}. The definitions of {\it core} and {\it compact core} are the same as in \citet{delaRosa2016}, i.e., {\it core} is the bulge component in bulge+disc decomposition of galaxies with all morphologies, while {\it compact core} is a core that fulfils the same compactness criterion used for massive compact galaxies at high redshift. The structural parameters of more than one million galaxies in SDSS DR7, such as their bulge and disc sizes, ellipticity, position and inclination angles, have been derived by performing bulge+disc decomposition \citep{Simard2011,Meert2015},  using the {\sf GIM2D} software package \citep{Simard2002}. We refer to the effective radius of the decomposed bulge component as $R_{{\rm e, core}}$. Based on the \citet{Simard2011} photometry catalog, \citet{Mendel2014} presented bulge, disc, and total stellar mass estimates of about 660,000 galaxies with spectroscopic redshifts and successful {\it ugriz} bulge+disc decomposition available, via fitting to broadband spectral energy distributions (SEDs). We refer to the estimated bulge stellar mass as $M_{\rm *,core}$. 

For high-redshift compact quiescent galaxy sample, we utilize the multiwavelength photometry in all five CANDELS/3D-HST fields \citep{Grogin2011,Koekemoer2011,Brammer2012,Momcheva2016}. We use the \citet{Skelton2014} catalogue, which includes photometric redshifts, rest-frame colors, and stellar population parameters derived by fitting the SED of each galaxy with a linear combination of seven templates via the {\sf EAZY} code \citep{Brammer2009}. The stellar masses are derived by fitting the SEDs using the {\sf FAST} code \citep{Kriek2009} based on the \citet{Bruzual2003} stellar population synthesis (SPS) models with the \citet{Chabrier2003} initial mass function (IMF) and solar metallicity. The effective radii of those galaxies are taken from the work of \citet{vanderWel2014}, which are measured with {\sf GALFIT} package \citep{Peng2002,Peng2010}. For our sample, we use the effective radii measured with the $HST$/WFC3 $H_{\rm F160W}$ images.

\subsection{Sample selection}

\subsubsection{Local galaxies with and without compact cores}\label{sub2.1}

Beginning with \citet{Mendel2014} catalogue, we select the local galaxy sample with compact cores following several steps. The basic idea is similar to that in \citet{delaRosa2016}. We summarize our selection steps as follows:
\begin{enumerate}
    \item Galaxies with uncertain structural decomposition, defined by using the parameter dBD$>1\sigma$, are firstly excluded.
    \item Those galaxies showing anomalous bulge+disc decomposition, tagged as Tpye 4 in the \citet{Mendel2014} catalog, are also excluded. 
    \item Disc inclination angle is set to $\leq$ 60 degrees, in order to minimize internal disc extinction on the bulge light.
    \item Bulge ellipticity  $e_{\rm core } $ is selected to be $< 0.6$, in order to avoid strong bars.
    \item Those galaxies with their effective radii of galaxies larger than 80 percent of the point spread function (PSF) are selected, in order to exclude the poorly resolved galaxies. 
    \item The redshift range is set to 0.02 $\leq z \leq$ 0.06, where the 3$\arcsec$ aperture of the SDSS spectroscopic fibre covers the physical size $\sim 0.6-1.8$ kpc, roughly corresponding to the effective radii of massive cQGs at $z\sim2$.
\end{enumerate}

Finally, we select massive galaxies with compact cores. We adopt the compactness criterion of the \citet{vD2015} definition:
\begin{equation}\label{equ:1}
\begin{array}{c}
  {\rm log}({M}_{*,{\rm core}}/{\rm M}_\odot) \geq 10.6, \\
  {\rm log}({R}_{\rm e,{\rm core}}/{\rm kpc}) < {\rm log}({M}_{*,{\rm core}}/{\rm M}_\odot) - 10.7
\end{array}
\end{equation}
We adopt the definition of circularized effective radius as $R_{\rm e} =  R_{\rm e,a} \times \sqrt{b/a}$, where $R_{\rm e,a}$ is the semi-major effective radius and $b/a$ is the axial ratio of the galaxy. By imposing these seven selection criteria, the final sample includes 182 objects. 

\begin{figure}
  \begin{center}
    \includegraphics[width=0.9\columnwidth]{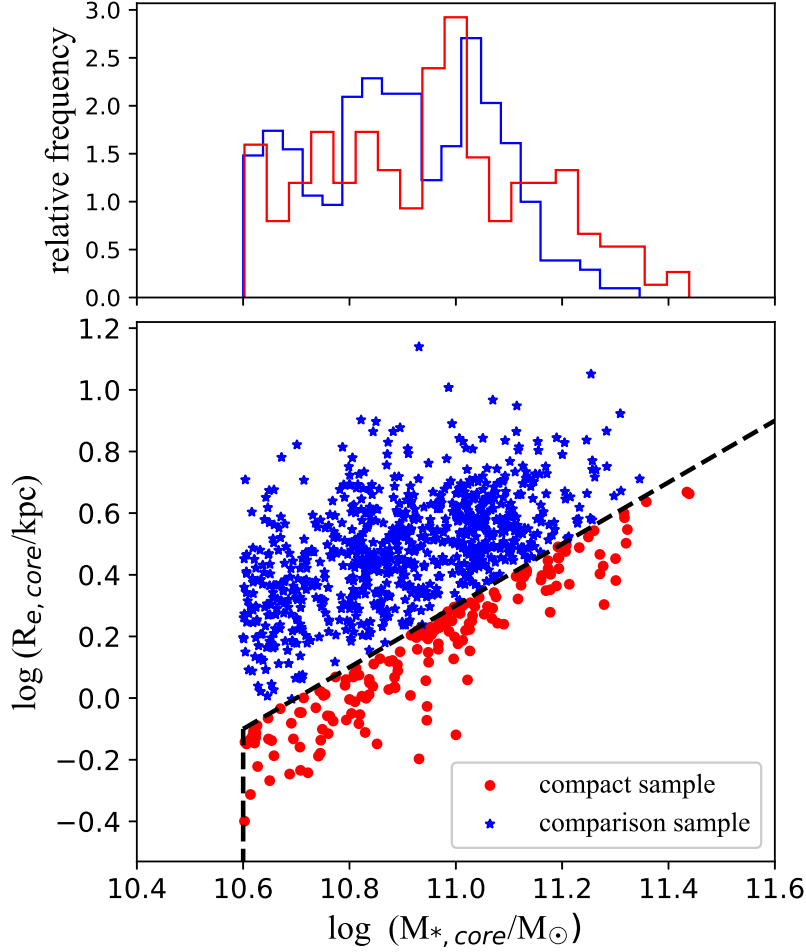}
  \end{center}
\caption{The mass-size distributions of local galaxies with compact cores and the galaxies in the comparison sample. In the lower panel, the blue symbols present the galaxies in the comparison sample, and the red points show the local galaxies with compact cores. The black dashed line marks the compactness criterion of the \citet{vD2015} definition. In the upper panel, we compare the stellar mass distributions of the compact sample (red line) and the comparison sample (blue line).}
\label{fig:1}
\end{figure}

We also build a comparison sample of galaxies without compact cores. For each galaxy with compact core, we select five comparison galaxies which fulfill the first six selection criteria presented before and have similar stellar masses ($\Delta {\rm log} M_* < 0.2$ dex). In the lower panel of Figure \ref{fig:1}, we plot the mass-size distributions of local galaxies with compact cores and the galaxies in the comparison sample. The black dashed line shows the compactness criterion of the \citet{vD2015} definition. In the upper panel, we compare the stellar mass distributions of these two samples.

\subsubsection{Massive compact quiescent galaxies at $1<z<3$}

As we propose to test the possible evolutionary connection between the local massive galaxies with compact cores and the high-$z$ cQGs, we need to construct the massive cQG sample at $1<z<3$. We follow the similar selection criterion presented in \citet{Fang2015} and \citet{Lu2019} to select the high-$z$ cQGs. Firstly, we select massive quiescent galaxies with log$({M}_{*}/{\rm M_\odot})\geq 10.6$ at $1<z<3$ by using the rest-frame $U-V$ and $V-J$ colors \citep{Wuyts2007,Williams2010}: 
\begin{equation}
\begin{array}{c}
V-J<1.5,\\
U-V>1.3,\\
U-V>0.8\times(V-J)+0.7
\end{array}
\label{equ:2}  
\end{equation}
Secondly, we select those massive quiescent galaxies utilizing the compactness criterion of the \citet{vD2015} definition.   Finally, we obtain a sample with 483 massive cQGs at $1<z<3$.

\section{Stellar population analysis} \label{sec:method}

Using the full-spectrum fitting code {\sf STARLIGHT} \citep{Fernandes2005} \footnote{https://www.starlight.ufsc.br}, which provides a fit to both galaxy continuum and spectral features, we fit the SDSS spectra of our local galaxy sample with compact cores and the comparison sample to derive their stellar population properties. The bases of {\sf STARLIGHT} are single stellar population (SSP) templates with a grid of ages and metallicities. To use {\sf STARLIGHT}, we  create a spectral library containing 45 SSPs from BC03 \citep{Bruzual2003} theoretical models, with 15 different ages ranging from 1 Myr to 13 Gyr and 3 different metallicities $Z = 0.004, 0.02, 0.05$. We adopt \citet{Chabrier2003} IMF and Galactic extinction law \citep{Cardelli1989} with R$_V$ = 3.1.

\begin{figure}
  \begin{center}
    \includegraphics[width=\columnwidth]{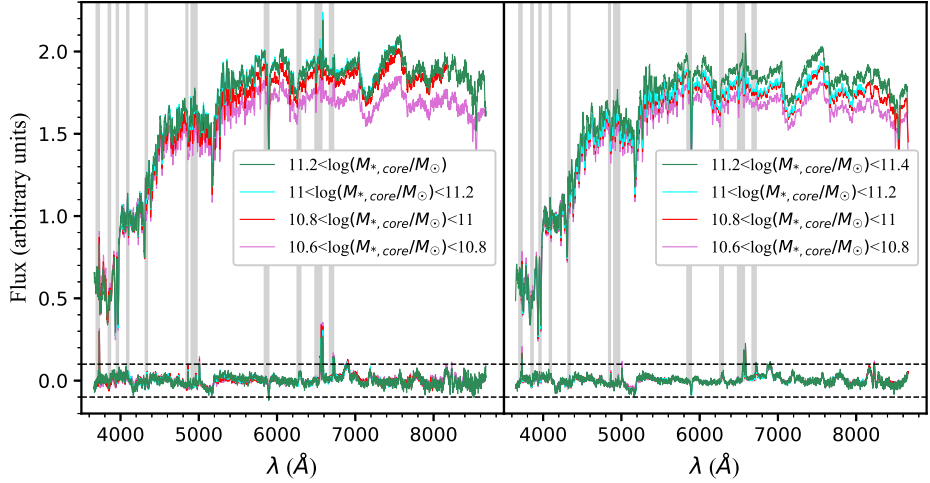}
  \end{center}
\caption{SDSS median stacked spectra for the samples of local massive galaxies with compact cores (left panel) and the comparison sample (right panel). The bottom of these two panels show the residual spectra (defined as the differences between the median stacked spectra and the {\sf STARLIGHT} best-fit spectra, see also Figure \ref{fig:3}). The gray shaded regions are the masked spectral regions. The median stacked spectra of four different mass bins are color-coded. }\label{fig:2}
\end{figure}

In order to increase the signal-to-noise (S/N) of the input spectra, we work on
the median stacked spectra derived for four mass bins: 10.6 $\leq {\rm log}({ M}_{*,{\rm core}}/{\rm M}_\odot) <$ 10.8, 10.8 $\leq {\rm log}({M}_{*,{\rm core}}/{\rm M}_\odot) < 11.0$, $11.0 \leq {\rm log}({M}_{*,{\rm core}}/{\rm M}_\odot) < 11.2$ and ${\rm log}({M}_{*,{\rm core}}/{\rm M}_\odot) \geq 11.2$. We obtain the median stacked spectra  following the same method presented in the work of \citet{Citro2016}. Here we summarize the main steps. Firstly, we shift each individual spectrum to rest-frame. Then we normalize each spectrum at rest-frame 5000\AA, where no strong absorption features are presented. Thirdly, we compute the median flux by interpolating the common wavelengths. Wavelength regions of some strong emission lines have been masked out (see  gray shaded regions in Figure \ref{fig:2}). We define the median stacked flux error as MAD/$\sqrt{N}$, where MAD\footnote{MAD=1.48$\times$median($|$X$_i-median(X_i$)$|$), the details can be found in \citet{Hoaglin1983}. } is the median absolute deviation and $N$ is the number of objects at each wavelength. 

\section{RESULTS AND DISCUSSION}\label{sec:result}

\begin{figure}
  \begin{center}
    \includegraphics[width=\columnwidth]{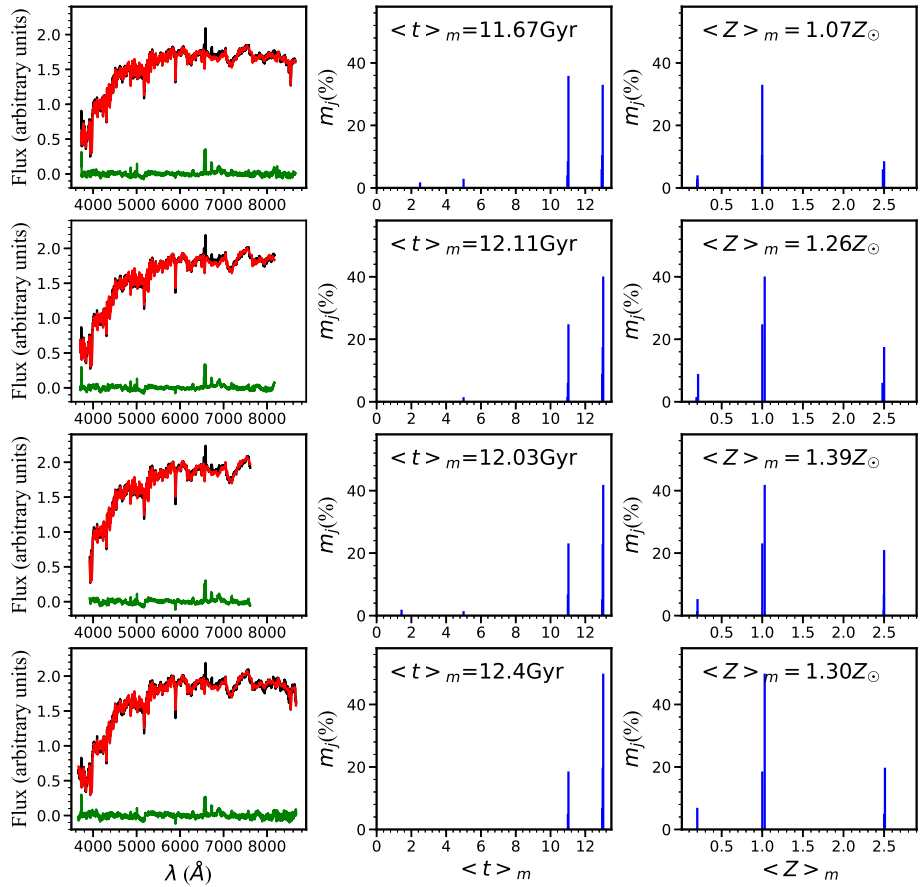}
  \end{center}
\caption{Results of {\sf STARLIGHT} fitting on the stacked spectra of local massive galaxies with compact cores.  We plot the results in one row for each mass bin with increasing mass from top to bottom. {\it Left}:  the {\sf STARLIGHT} best-fit spectra (red line), the stacked spectra (black line) and the residual spectra (green line). {\it Middle}: the distribution of mass-weighted ages. {\it Right}: the distribution of mass-weighted metallicities.}
\label{fig:3}
\end{figure}

\begin{figure}
  \begin{center}
    \includegraphics[width=\columnwidth]{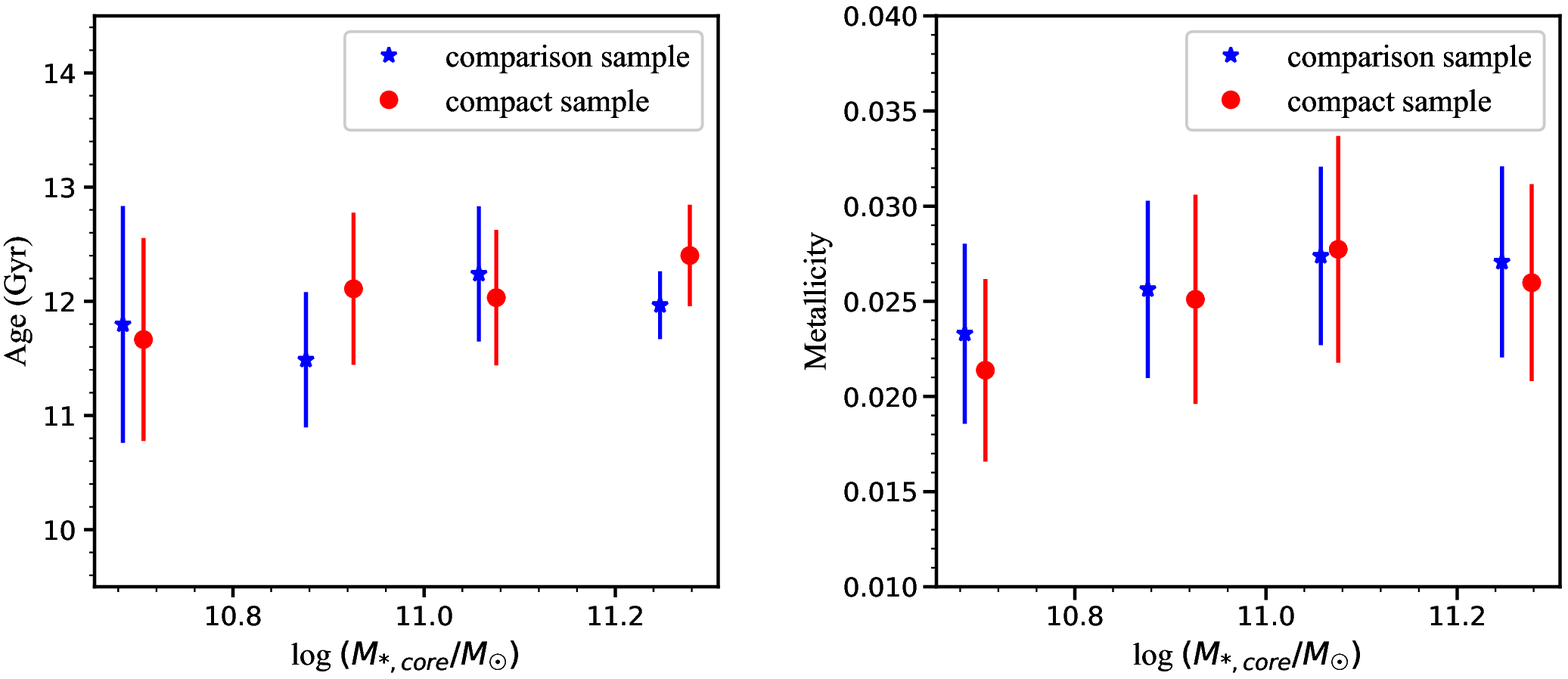}
  \end{center}
\caption{Mass-weighted ages (left panel) and metallicities (right panel) of local massive galaxies with compact cores (red) and the comparison sample (blue). }
\label{fig:4}
\end{figure}

\subsection{Ages and Metallicities}

By using {\sf STARLIGHT} fitting on the stacked spectra, we derive the mass-weighted ages and metallicities of both the local massive galaxy sample with compact cores and the comparison sample (see Figure \ref{fig:4}). The derived mass-weighted ages of local massive galaxies with compact cores vary from $\sim11.6$ to $\sim12.4$ Gyr, which are consistent with previous works at similar redshifts \citep{Thomas2010,Conroy2014,Citro2016}. Our results suggest that these compact cores have been formed at $z>2$. There is a weak evolutionary trend, with mass-weighted ages increasing systematically with stellar masses. This trend is in agreement with the downsizing scenario of galaxy evolution \citep{Cowie1996,Perez-Gonzalez2008,Fontanot2009}. We also find that metallicities correlate with stellar masses, well known as mass-metallicity relation \citep{Tremonti2004}. The relation gets flattened at the high mass end as shown in the previous works \citep{Tremonti2004,Gallazzi2005}. Compared to local massive galaxies with compact cores, those galaxies without compact cores show the age varying from $\sim11.5$ to $\sim12.2$ Gyr and the metallicity from 0.023 to 0.028. Considering the large uncertainties, we suggest that there is no obvious difference of ages and metallicities between local massive galaxy sample with compact cores and the comparison sample. This suggests that the massive cores of local galaxies are formed early, regardless of their compactness. 

\subsection{Local galaxies with compact cores as the possible descendants of high-$z$ cQGs}

\begin{figure}
  \begin{center}
    \includegraphics[width=\columnwidth]{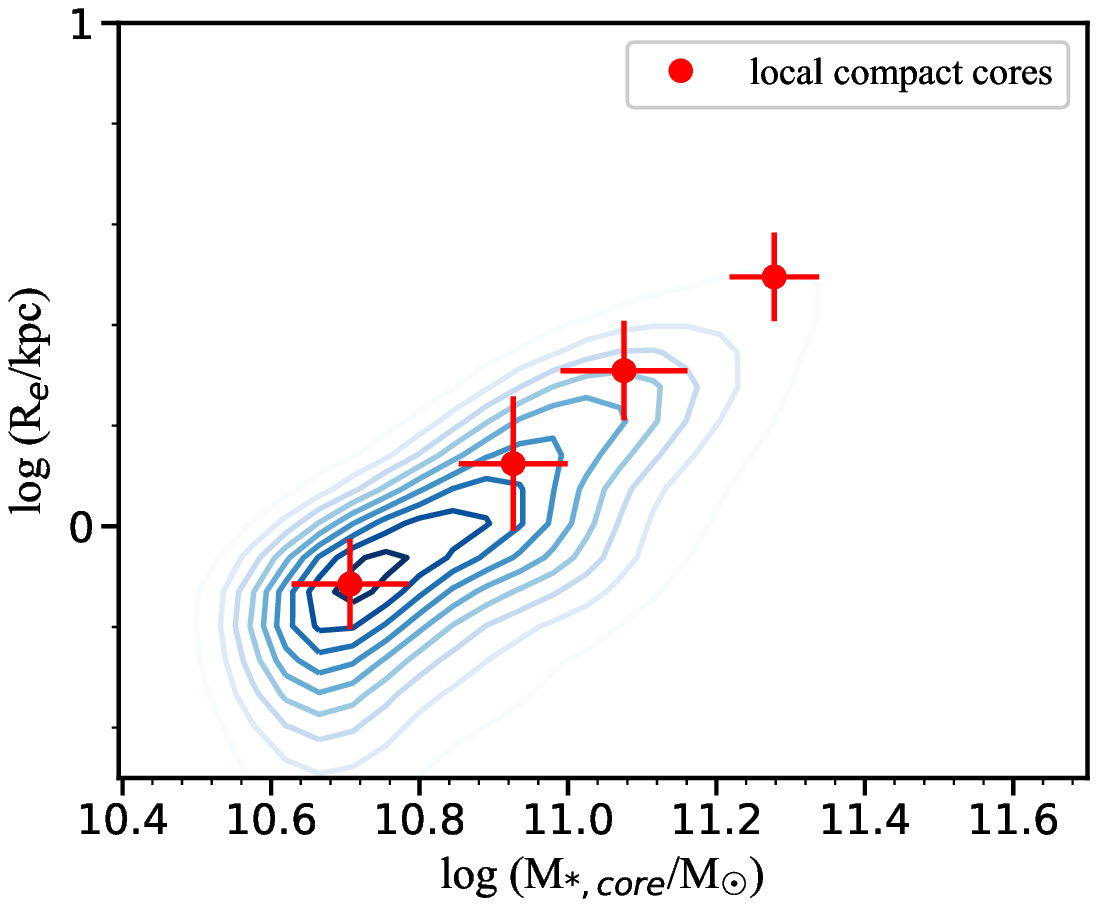}
  \end{center}
\caption{Mass-size distributions of local compact cores (red points) and cQGs at $1<z<3$ selected from CANDLES/3D-HST fields (color-coded contours). For cQGs at $1<z<3$, we refer to the y-axis $R_{\rm e}$ as the effective radius of the entire galaxy. For local compact cores (red points), we refer to the y-axis $R_{\rm e}$ as the effective radius of the bulge component $R_{\rm e,core}$.}
\label{fig:5}
\end{figure}

\begin{figure}
  \begin{center}
    \includegraphics[width=\columnwidth]{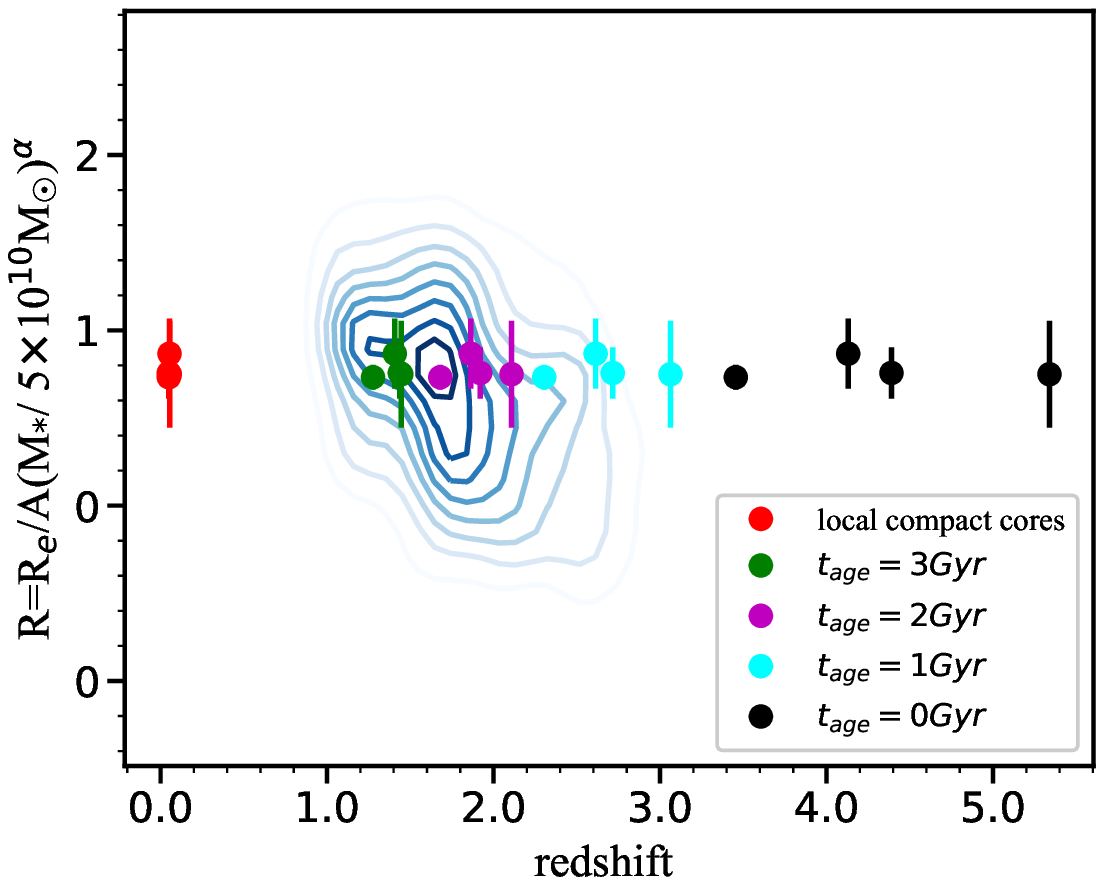}
  \end{center}
\caption{The normalized radii of local compact cores with different ages (color-coded points) and high-$z$ cQGs (color-coded contours) as a function of redshift. The normalized radius has been defined as $R_e/AM^{\alpha}$, where $M = M_*/5\times10^{10}{\rm M}_\odot$. The slope $\alpha$ is fixed to 1.0. The black, cyan, purple and green points mark the derived star formation histories of local compact cores when they are just formed, 1-Gyr-old, 2-Gyr-old and 3-Gyr-old, respectively. }
\label{fig:6}
\end{figure}

The main motivation of this paper is to test the possibility of local massive galaxies with compact cores as the potential descendants of massive compact galaxies at $1<z<3$. In Figure \ref{fig:5}, we plot the mass-size distributions of local compact cores and cQGs at $1<z<3$. For local compact cores, we use stellar masses and effective radii of compact cores, while we use those of the entire galaxies for cQGs at $1<z<3$. At the given stellar masses, local compact cores show the similar sizes to those of high-$z$ cQGs. In particular, mass-size relations of local compact cores and high-$z$ cQGs show the similar slope $\alpha\sim 1.0$ \citep{delaRosa2016}, where the definition of $\alpha$ follows $ R_e\sim M_*^\alpha$. We notice that the value of $\alpha\sim 1.0$ is the observational result for high-$z$ cQGs, while it is driven by the choice of our compactness criterion of \citet{vD2015}. For local massive ETGs, the slope $\alpha$ is different, varying from $\sim0.5$ to $\sim0.8$ \citep{Shen2003,Semyeong2017}. 

In Figure \ref{fig:6}, we plot the normalized radii of local compact cores with different ages and high-$z$ cQGs as a function of redshift. The normalized radius has been defined as $R_e/AM^{\alpha}$ \citep[e.g.,][]{vanderWel2014}, where $M =M_{*}/5\times10^{10}{\rm M}_\odot$ and $\alpha = 1.0$. With the derived mass-weighted ages, the formation redshifts of local compact cores (see black symbols in Figure \ref{fig:6}) have been suggested to be at $3<z<6$. Considering the derived stellar ages of cQGs at $1<z<2$ ranging from $\sim$1 to 3 Gyr \citep{Damjanov2009,Belli2015}, the formation redshifts of high-$z$ cQGs will be in the range of $2<z<6$, which is widely consistent with those of local compact cores. Assuming that local compact cores will evolve passively and have no size evolution after their formation, their normalized radii will keep constant. Cyan, purple and green points represent stellar ages of 1 Gyr, 2 Gyr and 3 Gyr after the formation of local compact cores. The normalized radii of local compact cores with stellar ages of 1$\sim$3 Gyr matches well with the observed values of cQGs at $1<z<3$, which indicates that local compact cores are possible descendants of high-$z$ cQGs. 

 \subsection{Morphology}
 
 If local massive galaxies with compact cores are possible descendants of high-$z$ cQGs, a relevant question is what kind of morphological types those galaxies will have. Following the same method as presented in \citet{delaRosa2016}, we classify local massive galaxies with compact cores into four morphology classes: Elliptical (Ell), lenticular (S0), Sab and Scd. According to the automated morphological classification based on support vector machines provided by \citet{Huertas-Company2011}, the probability of each galaxy morphology type has been given. With a simple linear model, these probabilities can be converted into the T-type classification scheme \citep{Meert2015}. By using the definition of the galaxy morphology classes in terms of T-type \citep{Meert2015}, each local galaxy with compact core can be flagged with one of four morphology classes. 
 
 In Table \ref{tab:3}, we present the morphology classification result of local massive galaxies with compact cores. Over 80\% of them host in elliptical and lenticular galaxies. However, there is still a non-negligible fraction ($\sim15\%$) of compact cores hosting in the early-type spiral galaxies (Sab). Only a few ($<3\%$) of them host in the late-type spiral galaxies (Scd). If the scenario that the high-$z$ cQGs survive as compact cores in the local universe is correct, there would be multiple possible evolutionary paths for high-$z$ cQGs. By building an extended wing according to dry minor merger \citep{Hopkins2009a,Naab2009,Patel2013}, most of high-$z$ cQGs will evolve into local massive ETGs with ``normal sizes", which match the local mass-size relation \citep{Shen2003}. Some of them build a substantial stellar/gas disc according to the late-time gas accretion and sustaining star formation,  and finally grow up to spiral galaxies.  
 
 \begin{table}
\begin{center}
\caption[]{ Morphology Classes Of Local Massive Galaxies With Compact Cores}\label{tab:3}


 \begin{tabular}{lcr}
  \hline\noalign{\smallskip}
Class & Definition  & Fraction \\
  \hline\noalign{\smallskip}
Elliptical  &   T-type $\leq -3$ &  39.6$\%$      \\
S0  &  $-3  < $T-type$\leq$ 0.5   &  42.3$\%$     \\
Sab &  0.5 $ < $T-type$\leq$ 4  &   15.4$\%$    \\
Scd &   T-type$>$ 4   &   2.7$\%$   \\
  \noalign{\smallskip}\hline
\end{tabular}
\end{center}
\end{table}

\section{SUMMARY AND CONCLUSIONS}\label{sec:sum}

Recent discoveries of high-$z$ cQGs provide a testbed for the different scenarios of massive galaxy evolution. Most studies in the literature proposed different evolutionary mechanisms under the assumption that the high-$z$ cQGs will evolve into the local ETGs. However, another potential evolutionary scenario has been recently proposed that the high-$z$ cQGs can survive as compact cores hosting in local massive galaxies with different morphology classes.  In this paper, we try to test this scenario by exploring the star formation histories of local massive compact cores according to their spectral and morphological analysis. We build a sample of 182 massive galaxies with compact cores ($M_{*,{\rm core}} > 10^{10.6} {\rm M}_\odot$) at $0.02 \leq  z \leq 0.06$ from SDSS DR7 spectroscopic catalogue. The narrow redshift range is selected to ensure the observed spectra coming from the central $\sim 1$ kpc region, which is about the average effective radius size of high-$z$ cQGs.

The sample is divided into four bins with increasing core masses and their median stacked spectra have been obtained. {\sf STARLIGHT} package is used to analyze the median stacked spectra and derive the stellar ages and metallicities. Our main results show that local compact cores have old ages with the average age of about 12 Gyr, indicating their early formation at $z > 3$, which is consistent with the formation redshifts of cQGs at $1<z<3$. Assuming that compact cores will evolve passively and have no size evolution after their formation, they will have the similar stellar ages (1-3 Gyr) and compact sizes as those of observed cQGs at $1<z<3$. Combined with the results presented in \citet{delaRosa2016}, who found that local compact cores have the similar structure and abundance as those of high-$z$ cQGs, our study supports the evolutionary scenario that high-$z$ cQGs survive as compact cores and embedded in the center of local massive galaxies. Morphological study of local galaxies with compact cores suggests that there would be multiple possible evolutionary paths for high-$z$ cQGs: most of high-$z$ cQGs will evolve into local massive ETGs by dry minor merger, while some of them build a substantial stellar/gas disc according to the late-time gas accretion and sustaining star formation,  and finally grow up to spiral galaxies.  

\begin{acknowledgements}
We thank the anonymous referee for constructive comments and suggestions. This work is supported by National Key R\&D Program of China (No. 2017YFA0402703). LF acknowledges the support from the National Natural Science Foundation of China (NSFC, Grant Nos. 11822303 and 11773020) and Shandong Provincial Natural Science Foundation, China (ZR2017QA001, JQ201801). YG thanks Dr. Z. Pan for the valuable discussion. 

Funding for the SDSS and SDSS-II has been provided by the Alfred P. Sloan Foundation, the Participating Institutions, the National Science Foundation, the U.S. Department of Energy, the National Aeronautics and Space Administration, the Japanese Monbukagakusho, the Max Planck Society, and the Higher Education Funding Council for England. The SDSS Web Site is {http://www.sdss.org/}.

The SDSS is managed by the Astrophysical Research Consortium for the Participating Institutions. The Participating Institutions are the American Museum of Natural History, Astrophysical Institute Potsdam, University of Basel, University of Cambridge, Case Western Reserve University, University of Chicago, Drexel University, Fermilab, the Institute for Advanced Study, the Japan Participation Group, Johns Hopkins University, the Joint Institute for Nuclear Astrophysics, the Kavli Institute for Particle Astrophysics and Cosmology, the Korean Scientist Group, the Chinese Academy of Sciences (LAMOST), Los Alamos National Laboratory, the Max-Planck-Institute for Astronomy (MPIA), the Max-Planck-Institute for Astrophysics (MPA), New Mexico State University, Ohio State University, University of Pittsburgh, University of Portsmouth, Princeton University, the United States Naval Observatory, and the University of Washington.
\end{acknowledgements}

\label{lastpage}

\end{document}